\documentclass[a4paper,twocolumn]{emulateapj}
\usepackage{amstext}
\usepackage{apjfonts}
\usepackage{amsmath}
\usepackage{natbib}
\shortauthors{Naiman et al.}

\begin{document}

\title{Gas Accretion by Star Clusters and the Formation of
  Ultraluminous X-ray Sources from Cusps of Compact Remnants}

\author{J. P. Naiman\altaffilmark{1}, Enrico
  Ramirez-Ruiz\altaffilmark{1} and Douglas N. C. Lin\altaffilmark{1}}
\altaffiltext{1}{Department of Astronomy and Astrophysics, University
  of California, Santa Cruz, CA 95064; jnaiman@ucolick.org}

\begin{abstract}
Here we show that the overabundance of ultra-luminous, compact X-ray
sources (ULXs) associated with moderately young clusters in
interacting galaxies such as the Antennae and Cartwheel can be given
an alternative explanation that does not involve the presence of
intermediate mass black holes (IMBHs). We argue that gas density
within these systems is enhanced by the collective potential of the
cluster prior to being accreted onto the individual cluster members
and, as a result, the aggregate X-ray luminosity arising from the
neutron star cluster members can exceed $>10^{39}\;{\rm erg s^{-1}}$.
Various observational tests to distinguish between IMBHs and accreting
neutron star cusps are discussed.
\end{abstract}


\section{Introduction}
Over the years, the existence of two distinct populations of black
holes has been established beyond a reasonable doubt.  Supermassive
black holes, $M > 10^6 \, M_\sun$, are inferred in many galactic
centers \citep{1995ARAA..33..581K,1998AJ....115.2285M}, while stellar
mass black holes, $M \sim 1-10 \, M_\sun$, have been identified by
their interaction with companion stars \citep{2006csxs.book..157M}.
The situation at intermediate masses, $M \sim 10^2 - 10^5 M_\sun$, is
still uncertain despite recent evidence for mass concentrations within
the central regions of some globular clusters
\citep{2005ApJ...634.1093G,2007ApJ...661L.151U,2008ApJ...676.1008N}.
This evidence remains controversial, partly because the velocity
dispersion profiles can be reproduced without invoking the presence of
an intermediate mass black hole (IMBH)
\citep{baum1, baum2,2009arXiv0905.0627A}.

Recently, some evidence has arisen for the presence of IMBHs in
moderately young star clusters, where ultra-luminous, compact X-ray
sources (ULXs) have been preferentially found to occur
\citep{2001ApJ...554.1035F,2008AIPC.1010..357T}.  Their high
luminosities have been interpreted as imprints of 
IMBHs \citep{2004Natur.428..724P}, rather than
binaries containing a normal stellar mass black hole
\citep{2006ApJS..166..211Z}. In this {\it Letter}, we present an
alternative explanation for the overabundance of ULXs associated with
young clusters in galaxies such as the Antennae and Cartwheel.  In
this new paradigm, the accretion of gas by the collective star cluster
potential moving through the merging medium is strongly enhanced
relative to the individual rates and, as a result, the aggregate X-ray
luminosity arising from the neutron star cluster members can
exceed $>10^{39}\;{\rm erg s^{-1}}$.  Much of the effort herein will
be dedicated to understanding the conditions by which the collective
potential of a star cluster is able to accrete gas with highly
enhanced rates and its effect on the integrated accretion luminosity
of the neutron star cluster members.  The conditions found in systems
such as the Antennae galaxy, as we will argue, are favorable for this
type of mechanism to operate effectively and produce an overabundance
of ULXs.

\section{ULX Cusps from Compact Stellar Cluster Members}
ULXs are seen in the star clusters of merging galaxies, such as
the Antennae and the Cartwheel \citep{2008AIPC.1010..357T,
  2006ApJS..166..211Z}.  These sources are compact in nature and in
general associated with super star clusters (SSCs) - young, compact,
massive clusters of stars \citep{2002ApJ...577..710Z}. Many of these
sources have luminosities $\ge 10^{39} \, {\rm erg \, s^{-1}}$, which
suggest that they could be IMBHs rather than binaries containing a
normal stellar mass black hole.

A compact star of mass $M_\ast$, moving with relative velocity $v$
through a gas of ambient density $\rho$ and sound speed $c_s$,
nominally accretes at the Bondi-Hoyle-Lyttleton rate: $\dot{M}\eqsim
4\pi (GM_\ast)^2\rho (v^2+c_s^2)^{-3/2}$
\citep{2004NewAR..48..843E}. For a NS of mass $M_\ast = M_{\rm NS}=1.4 M_\odot$
and radius $R_{\rm NS}=10\;{\rm km}$, the corresponding X-ray
luminosity is given by
\begin{equation}
L_X = \epsilon G M_{\rm NS} \dot{M} R_{\rm NS}^{-1}=10^{32}n\epsilon
\left({V \over 10{\rm \;km/s}}\right)^{-3}\;{\rm erg/s},
\end{equation}
where $\epsilon\leq1$ is the efficiency for converting gravitational
energy into X-ray radiation, $n=\rho/m_p$ is the hydrogen number
density in units of cm$^{-3}$ and $V=(v^2+c_s^2)^{1/2}$. The
integrated X-ray accretion luminosity of $N_{\rm NS}$ neutron star cluster
members is then given by $L_X=10^{36}(N_{\rm NS}/10^4) n \epsilon(V/10{\rm
  \;km/s})^{-3}\;{\rm erg/s}$.

In order for an aggregated accretion model to successfully describe
ULXs, the predicted X-ray luminosity must naturally span the range of
observed luminosities. This requires that the resulting speed $V$ not
be too large but more importantly that the external density be
relatively high. Direct observational searches for cluster gas  in the
form of molecular, neutral, and ionized hydrogen have yielded
non-detections, implying upper limits on the total gas content in the
range of 0.1-10 $M_\odot$ \citep{1995MNRAS.273..632S}.
In a search for ionized gas \citet{1996ApJ...462..231K}
found upper limits of 0.1 $M_\odot$ within about one core radius for
the clusters, implying $n_{\rm H+} \leq 50\;{\rm cm^{-3}}$.

A simple argument can be made to determine a lower limit to the
density of the gas in the cores of globular clusters (GCs)
 in the absence of
gas retention \citep{2001ApJ...550..172P}.
Suppose that the inner core of a GC contains $N_\ast=10^2N_{\ast,2}$
red giant stars, and so their mean separation is $r_\perp= 6.4 \times
10^{16} N_{\ast,2}^{-1/3} r_{\rm c,-1}\;{\rm cm}$, where $r_{\rm
  c}=0.1r_{\rm c,-1}$ pc. A lower limit on the wind density can be 
made by assuming that the wind of each of the stellar member extends only to its 
closest neighbors.  In this approximation, $n > n_{\rm w} =80
N_{\ast,2}^{2/3}r_{\rm c,-1}^{-2} v_{\rm w,1}^{-1} \dot{M}_{\rm w, -7}
\,{\rm cm}^{-3}$, where $v_{w}= 10 v_{\rm w,1}$ km s$^{-1}$
and $\dot{M}_{\rm w}=10^{-7}\dot{M}_{\rm w,-7}$ M$_{\odot}$ yr$^{-1}$
are the velocity and mass loss rate of the stellar core members. When
the cluster gravity and the interaction between stellar winds is taken
into account, we suspect that the gas density can be larger than this
value.

For clusters that are moving through a relatively dense medium, as in
Antennae galaxy for which CO measurements give $n\sim 10^{3}$
cm$^{-3}$ \citep{2003ApJ...588..243Z}, the collective external mass
accretion is likely to shape the luminosity function for the accreting
distribution of neutron stars.  A density of 10$^{3}$ cm$^{-3}$,
however, gives an aggregate neutron star luminosity of about
$L_X=10^{38}(N_{\rm NS}/10^4)$ $(\epsilon/0.1)(V/10{\rm \;km/s})^{-3}$
erg/s, which is not high enough to explain ULXs
\citep{2004ApJ...601L.171K,2007Natur.445..183M,2001ApJ...550..172P}.
If, however, the collective potential of the cluster was able to
significantly increase the surrounding gas density prior to being
accreted onto the individual neutron star members, the aggregate X-ray
luminosity could exceed $10^{39}$ erg/s. It is to this problem that we
now turn our attention.

\section{The Cluster Model and Numerical Method}
To test the gas density enhancement efficiency of a star cluster, we
simulated a cluster potential moving through the merging galaxy medium
at various typical velocities using FLASH, a parallel, adaptive mesh
refinement hydrodynamics code. This scheme, and tests of the code are
described in \citet{2000ApJS..131..273F}.  All star clusters are
modeled here with a Plummer potential:
\begin{equation}
\Phi = \frac{G M_{\rm c}}{(r^2 + r_{\rm c}^2)^{1/2}}  
\end{equation}
Here, $M_{\rm c}$ is the total cluster mass, taken to be $3.5 \times
10^{5} M_\odot$ \citep{1999AAS...195.4715Z,2007ApJ...668..168G}.  We
use several typical SSC cluster core radii, $r_{\rm c} = 1, 2, 3$ pc
\citep{2007ApJ...663..844M}.  For comparison,
\citet{1999AJ....118.1551W} estimates the typical half-light radius of
the Antennae clusters to be $(4 \pm 1) \, {\rm pc}$.  The
core radius is expected to be significantly less than the half-light
radius.

Our main goal is to examine the ability of a potential to accrete gas
as a function of the relative speed of the potential through the gas,
and the gas temperature.  Our star cluster, here modeled as a Plummer
potential, has been therefore set in motion through an initially
uniform medium. The speed of sound far away from the cluster is taken
to be $c_{\rm s}\sim 10 \, {\rm km \, s^{-1}}$, which is consistent
with the inferred intracluster medium temperature $\sim 10^{4} \,
 K$ in the Antennae galaxy \citep{2007ApJ...668..168G}.  Based
on observations of several cluster knots in the Antennae galaxy, which
indicate intracluster velocity dispersions of order $10 \, {\rm km \,
  s^{-1}}$ \citep{2005AJ....130.2104W}, the initial Mach number of the
cluster relative to the gas is varied between $\mu_\infty =v_\infty/c_s = 0.5$ to
$4.0$.  Here, $v_\infty$ and $c_s$ are the velocity of the medium and the sound 
speed at infinity, respectively. The gas within the dense medium has a temperature selected to
give the desired value of $c_{\rm s}$ and a density $\rho_\infty=
10^{-21}\, {\rm g \, cm^{-3}}$, chosen to match the intracluster
densities as derived from CO measurements \citep{2003ApJ...588..243Z}.

The effects of self gravity of the gas are ignored. This is adequate
for most of our models, for which the accreted mass is less than the
mass responsible for the potential.  To improve the controlled nature
of the models, we do not explicitly include radiative heating or
cooling. The gas, instead, evolves adiabatically. The effects of
radiative equilibrium are approximated by having the gas evolve with
an adiabatic constant $\gamma= 1.01$, giving nearly isothermal
behavior, which is consistent with the presence of a large quantity of
dust near these clusters as inferred from infrared observations
\citep{2005ApJ...635..280B}.  In cases where sufficient gas is
accreted for it to become self-shielded, cooling could decrease the
temperature of the gas significantly, potentially enhancing the
accretion rate beyond the values computed here.

We use inflow boundaries on one side of our rectangular grid to
simulate the cluster's motion through the ambient media.  We run our
simulations from initially uniform background density until a steady
density enhancement forms in the cluster center, which usually takes a
few 10-100 sound crossing times.  Several models were run longer to
test convergence and density enhancements were found to change only
slightly with longer run times.  We further tested convergence of our
models for several resolutions and domain sizes.  All tests produced
similar density enhancements to those shown here.  After hundreds of
sound crossing times, the flow is relatively stable, and does not
exhibit the ``flip-flop'' instability seen in two dimensional
simulations \citep{2009ApJ...700...95B}.

\section{Resulting Mass Density Profiles and ULX Cusps}
The accretion of ambient gas by moving bodies is a classical
problem. Many studies have been focused on the flow around compact
stars with a point mass potential.  Although clusters have much larger
masses than individual stars, their potential is relatively shallow
and the classical treatment derived for a point-mass potential is only
a fair approximation far from the cluster when $GM_{\rm c}/r_{\rm c}
\gg c_{\rm s}^{2}+v^{2}$. When $GM_{\rm c}/r_{\rm c} \lesssim c_{\rm
  s}^{2}+v^{2}$, the collective potential alters the local gas
properties before the gas is accreted onto the individual stars within
the cluster.

\begin{figure}
\centering\includegraphics[height=0.8\textwidth,
  angle=270]{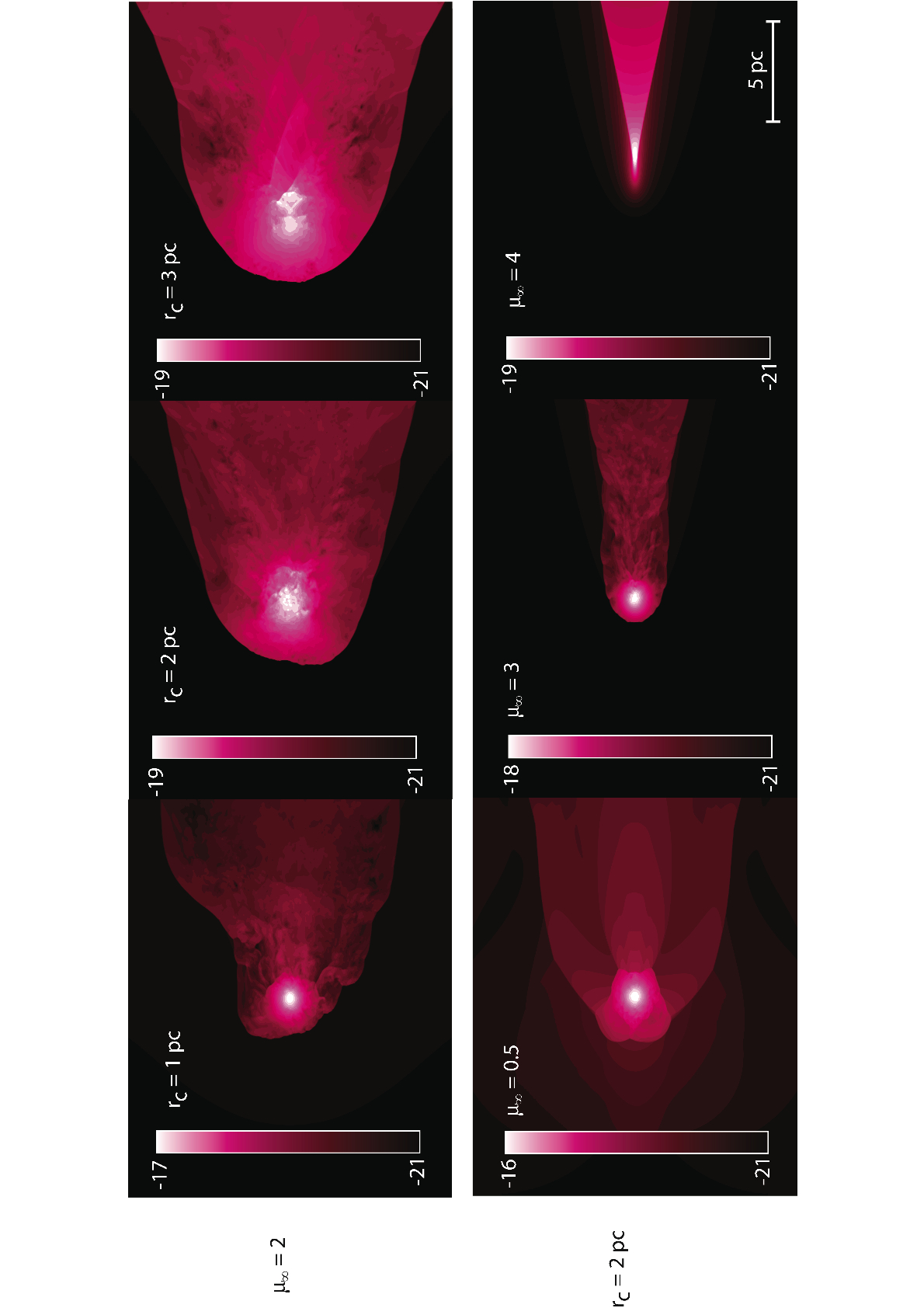}
\caption{The flow pattern around a star cluster set in motion through
  an initially uniform medium with varying core dimensions (${\rm
    r_c}$) and relative speeds ($\mu_\infty$).  Color bars show
  density cuts through the {\it{xy}}-plane in units of $\rm {log
    (g/cm^3)}$.}
\label{fig:fig1}
\end{figure}

\begin{figure}
\centering\includegraphics[height=0.8\textwidth, angle=270]{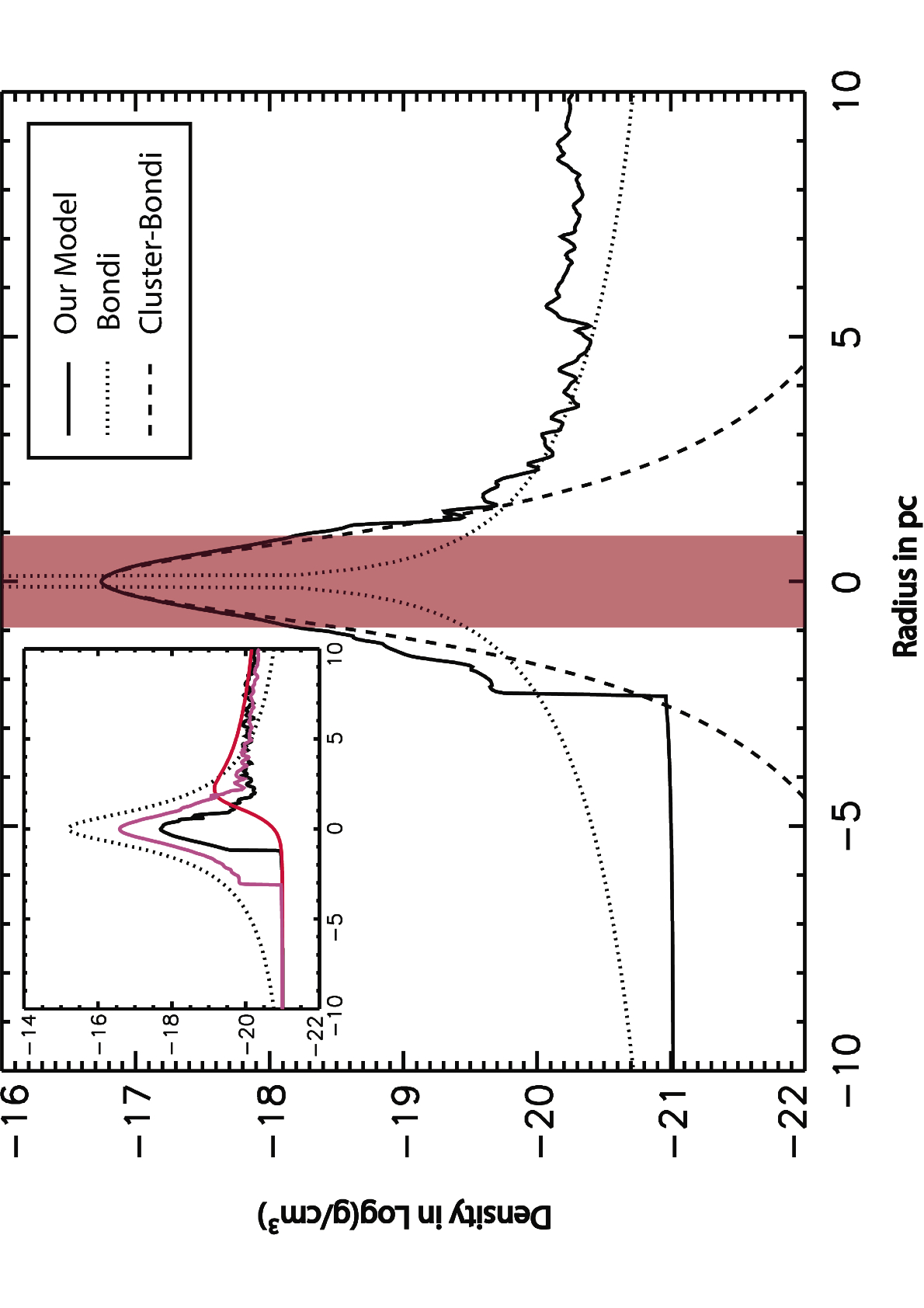}
\caption{Density profile for a model with core radius $r_{\rm c} = 1$
  pc and $\mu_\infty = 2.0$.  Solid black line shows a cut along the
  x-axis for the simulated cluster. The the dotted line is the Bondi
  solution for a stationary point mass, while the dashed line gives
  the renormalized Bondi-Hoyle-Lyttleton solution modified for a
  cluster potential as derived by \citet{2007ApJ...661..779L}.  These
  modified solutions provide a good description of the density profile
  for $\mu_\infty \leq 1$ but not its normalization. As expected,
  these analytical solutions fail to reproduce the simulated flow for
  high Mach numbers as shown in the inset panel. {\it{Inset:}} The
  dotted line is the unnormalized cluster-Bondi solution while the
  light pink, black and red lines show the numerical profiles for
  $\mu_\infty = 2.0$, $3.0$, and $4.0$, respectively.}
\label{fig:fig2}
\end{figure}

Figure \ref{fig:fig1} shows the resulting density profiles for star
clusters with $GM_{\rm c}/r_{\rm c} \sim c_{\rm s}^{2}+v^{2}$ for a
variety of core radius and relative motions with respect to the
external medium combinations.  For small $r_{\rm c}$, the potential
starts to resemble that of a point mass and, as a result, the density
enhancement in the central regions is very significant.  A density
enhancement is observed to persist as long as the sound speed or the
relative velocity of the ambient medium is greater than the central
velocity dispersion of the cluster.  The enhanced density profiles
within the cluster differ from the classical Bondi solution, and, for
low Mach numbers, are better described by the cluster-Bondi analytic
profiles derived by \citet{2007ApJ...661..779L} as depicted in Figure
\ref{fig:fig2}.  The profiles begin to deviate significantly from the
cluster-Bondi solutions for high Mach numbers (inset in Figure
\ref{fig:fig2}).

The flow pattern around a star cluster at large relative velocities is
multi-dimensional and complex (Figure \ref{fig:fig1}). In the frame of
the potential, the gas streamlines are bent towards the cluster
center. Some shall intersect the center, while others converge along a
line behind it. The convergence speed of the gas determines the
reduction in its velocity relative to the potential due to shocks, and
therefore whether or not the gas is accreted. In line with the
conventional treatment, clusters moving with respect to the
interstellar medium at increasing supersonic velocities will have
density enhancements that are progressively lower and significantly
more offset from the cluster's center.  In these cases the aggregate
mass accretion rate of the central neutron star is not significantly
increased and stars accrete gas as though they move through the
external medium independently. Because several gas knots in the
Antennae galaxy have velocities relative to the cluster of order the
sound speed of the ambient medium, gas within these cluster cores
would achieve high densities. Within this environment, accretion by
the individual cluster members will be enhanced greatly relative to
their rate of accretion directly from the ambient gas.

\subsection{X-ray Luminosities from Enhanced Accretion Rates}

Accretion and emission from the neutron star cluster strongly depends
on the radial distribution of both compact remnants and gas. In this
model, we calculate the expected X-ray emission from the neutron star
members in the cluster using two extreme examples for the radial
distribution of compact remnants. The first one is based on
Fokker-Planck models of a core collapsed (centrally condensed)
globular cluster \citep{1997ApJ...481..267D}, and the second simply
assumes that the neutron stars, containing 1\% of the total mass,
follow the radial stellar mass distribution. The absorption corrected
X-ray luminosities and characteristic emission frequencies of the
neutron star cusps are calculated assuming a neutral absorbing medium
with solar metallicity. Note that in this paper we consider neutron
stars to be magnetic field free. If their fields are strong enough,
the propeller effect may reduce the X-ray luminosity of neutron stars
\citep{1999ApJ...520..276M}.

\begin{figure}
\centering\includegraphics[width=0.8\textwidth]{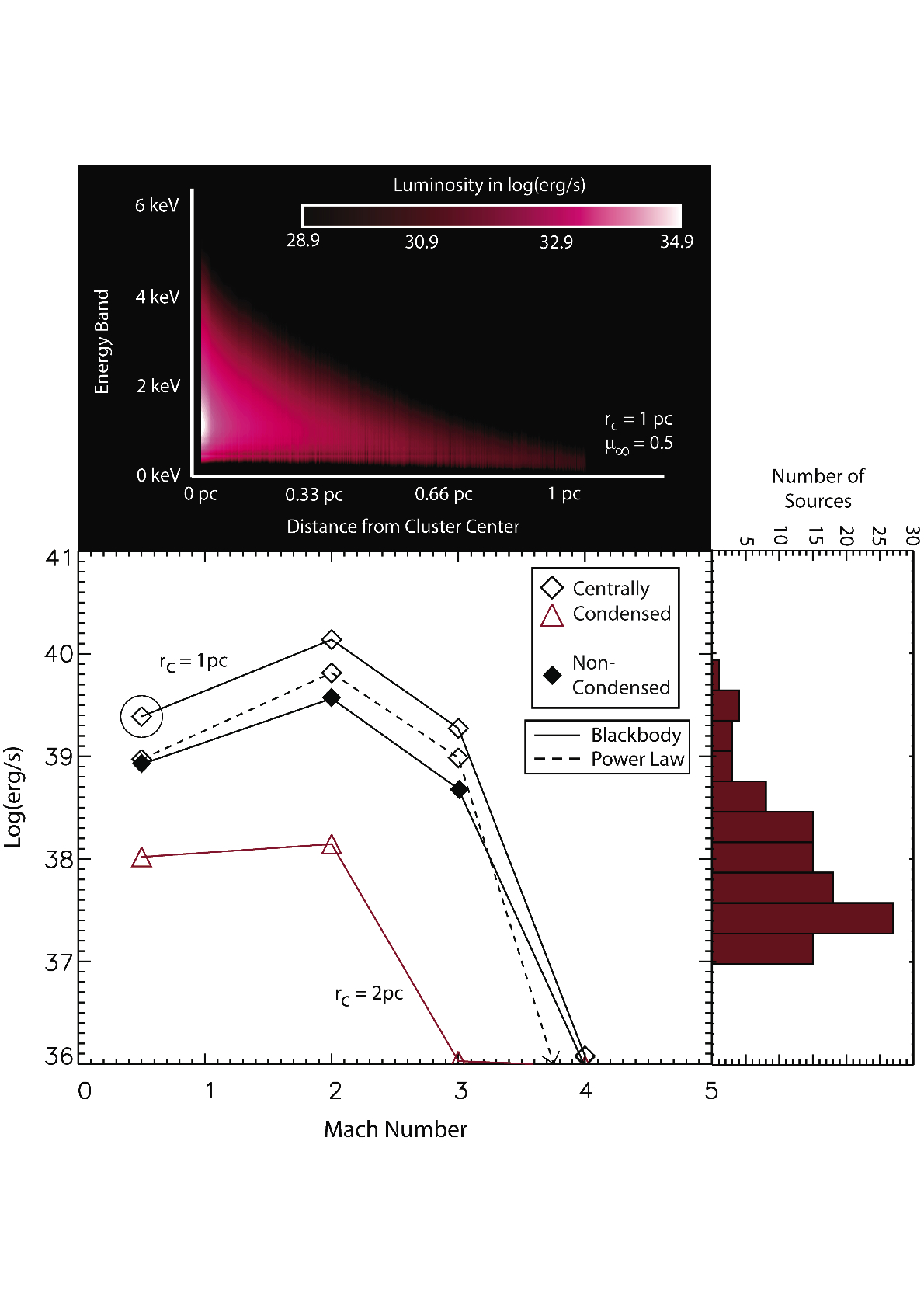}
\caption{X-ray luminosities from enhanced accretion
  rates. {\it{Center:}} The absorption corrected X-ray luminosities
  from an accreting neutron star members in a star cluster as a
  function of $\mu_\infty$ calculated using two extreme examples for
  the radial distribution of compact remnants.  Triangles are for
  $r_{\rm c} = 2 \, {\rm pc}$ while diamonds are for $r_{\rm c} = 1 \,
  {\rm pc}$. {\it Upper:} Spectral and luminosity decomposition as a
  function of distance from the cluster's center for a model with
  $\mu_\infty = 0.5$, $r_{\rm c} = 1 \, {\rm pc}$.  {\it{Right:}}
  Luminosity distribution of the X-ray sources in the Antennae galaxy
  \citep{2007ApJ...661..135Z}.}
\label{fig:fig3}
\end{figure}

Figure \ref{fig:fig3} shows the aggregate X-ray luminosity of the
accreting neutron star cluster as a function of the relative Mach
number for both centrally condensed and non-condensed compact remnant
distributions. The upper panel shows, for the centrally condensed
case, how the absorption corrected X-ray luminosities vary with both
photon energy and radial position within the cluster. As argued above,
clusters moving with increasing supersonic velocities will have
density enhancements that are progressively lower and significantly
more offset from the cluster's center. As a result, the aggregate
X-ray luminosity rapidly decreases with increasing Mach number
(although less sharply for more extended clusters). While the density
enhancement increases monotonically with decreasing relative velocity,
so does the corresponding photoionization absorption. These two
competing effects produce a maximum in the X-ray luminosity of the
cluster at about $\mu_\infty = 2$.  Most of the luminosity comes from
95\% (60\%) of all neutron stars in the condensed (non-condensed)
distributions with X-ray luminosities ranging between $2\times
10^{34}$ ($10^{33}$) and $4 \times 10^{34}$ ($2 \times 10^{34}$) ${\rm
  erg \, s^{-1}}$.  In this regime, the results depend weakly on the
assumed radiation spectra of the individual accreting members.

\section{Discussion}
Many studies have been focused on the flow around compact stars with a
point mass potential. Although clusters have much larger masses than
individual stars, their potential is relatively shallow. In this paper
we consider the efficiency of accretion in these cluster potentials,
and show that when the sound speed or the relative velocity of the
ambient medium is less than the central velocity dispersion of the
cluster, the collective potential alters the local gas flow before the
gas is accreted onto the individual stars within the
cluster. Accretion onto these dense stellar cores at the inferred rate
can lead to the onset of ULX sources.

While there are no stellar clusters observed in the Galactic disk
which bear these anticipated properties (the relative velocity of the
halo clusters to the interstellar medium is in the range of 100 km/s),
observations of several cluster knots in the Antennae indicate
intracluster relative velocities that comparable to the central
velocity dispersions \citep{2005AJ....130.2104W}.  Based on the
results of the current work, we show that accretion by individual
compact stars in the centers of such systems is enhanced greatly
relative to their rate of accretion directly from the ambient gas, and
conclude that this process may be relevant for explaining the origin
of ULX sources in these extraordinary clusters.  Illumination of
nearby gas clouds by these sources may also lead to reprocessed
infrared, optical and ultraviolet emission.  Finally, the sources may
leave trails of denser and likely hotter gas behind them as they
plough through the gas.

A way to distinguish between an IMBH \citep{2004Natur.428..724P} and
an accreting neutron star cusp is via time-dependent observations. The
emission from a relativistic region of a IMBH might vary on
time-scales of seconds.  The emission of a large number of
statistically independent black holes and neutron stars should be
considerably less variable: $\Delta t \leq r_{\rm c}/c_{\rm s}\sim
10^4$ yr. Observations find that a handful of the X-ray sources in the
Antennae galaxy are indeed variable albeit on timescales that are
larger than a few years \citep{2006ApJS..166..211Z}. Such variability
might be explained if only a moderate fraction of the compact stars
dominate the total luminosity. Such compact isolated accretors will
probably have unusual time-variability properties as their discs may
be much larger than the typical discs of X-ray binaries, and indeed
they are missing the perturbing influence of the secondary. On the
other hand, accretion disc feeding in these sources will be variable
itself, leading to variability on variety of time-scales.

Although accretion disk spectra are notoriously difficult to calculate
from first principles, an IMBH and a cluster core may also have
observably different spectra. It has been suggested that a cool
multicolor disk spectral component, might indicate the presence of an
IMBH \citep{2003ATel..212....1M}.  This is understood as
following. The larger mass of the BH accretor, the lower the
temperature of the inner edge of the disk, which scales as $T_{\rm i}
\propto (M_{\rm BH}\dot{M}/r_{\rm i})^{1/4} \propto
\dot{M}^{1/4}M_{\rm BH}^{-1/2}$ for a simple thin-disk model. Since
for our model, individual neutron star sources have $T_{\rm i}\sim 1$
keV (consistent with observations), then an IMBH might have $T_i \sim
30$ eV. Thus, an association of the ULXs with an IMBH, as opposed to a
accreting distribution of compact remnants, could be made on the basis
of a very soft observed spectral component.  We know of no reported
observations of such components.

\acknowledgements

We thank Jason Kalirai and Glenn van de Ven for useful discussions.
The software used in this work was in part developed by the
DOE-supported ASCI/Alliance Center for Astrophysical Thermonuclear
Flashes at the University of Chicago. Computations were performed on
the Pleaides UCSC computer cluster. This work is supported by NSF:
PHY-0503584 (JN and ER), NASA: NNX08AL41G (JN and DL) and The David
and Lucile Packard Foundation (ER).

\clearpage

\clearpage

\clearpage

\clearpage

%\bibliography{ms.bib}

\begin{thebibliography}{}

\bibitem[Kormendy 
\& Richstone(1995)]{1995ARAA..33..581K} Kormendy, J., \& Richstone, D.\ 1995, \araa, 33, 581 

\bibitem[Magorrian et al.(1998)]{1998AJ....115.2285M} Magorrian, J., et 
al.\ 1998, \aj, 115, 2285 


\bibitem[McClintock 
\& Remillard(2006)]{2006csxs.book..157M} McClintock, J.~E., \& Remillard, R.~A.\ 2006, Compact stellar X-ray sources, 157 


\bibitem[Gebhardt et al.(2005)]{2005ApJ...634.1093G} Gebhardt, K., Rich, 
R.~M., \& Ho, L.~C.\ 2005, \apj, 634, 1093 

\bibitem[Baumgardt et al.(2003)]{baum1} Baumgardt, H., Hut, 
P., Makino, J., McMillan, S., \& Portegies Zwart, S.\ 2003, \apjl, 582, L21 

\bibitem[Baumgardt et al.(2003)]{baum2} Baumgardt, H., 
Makino, J., Hut, P., McMillan, S., \& Portegies Zwart, S.\ 2003, \apjl, 589, L25 

\bibitem[Fabbiano et al.(2001)]{2001ApJ...554.1035F} Fabbiano, G., Zezas, 
A., \& Murray, S.~S.\ 2001, \apj, 554, 1035 

\bibitem[Zezas et al.(2006)]{2006ApJS..166..211Z} Zezas, A., Fabbiano, G., 
Baldi, A., Schweizer, F., King, A.~R., Ponman, T.~J., 
\& Rots, A.~H.\ 2006, \apjs, 166, 211 


\bibitem[Trinchieri et al.(2008)]{2008AIPC.1010..357T} Trinchieri, G., 
Wolter, A., 
\& Crivellari, E.\ 2008, A Population Explosion: The Nature \& Evolution of X-ray Binaries in Diverse Environments, 1010, 357 

\bibitem[Zezas et al.(2002)]{2002ApJ...577..710Z} Zezas, A., Fabbiano, G., 
Rots, A.~H., \& Murray, S.~S.\ 2002, \apj, 577, 710 


\bibitem[Edgar(2004)]{2004NewAR..48..843E} Edgar, R.\ 2004, New Astronomy 
Review, 48, 843 


\bibitem[Zhang 
\& Fall(1999)]{1999AAS...195.4715Z} Zhang, Q., \& Fall, S.~M.\ 1999, Bulletin of the American Astronomical Society, 31, 1443 

\bibitem[Gilbert 
\& Graham(2007)]{2007ApJ...668..168G} Gilbert, A.~M., \& Graham, J.~R.\ 2007, \apj, 668, 168 

\bibitem[McCrady 
\& Graham(2007)]{2007ApJ...663..844M} McCrady, N., \& Graham, J.~R.\ 2007, \apj, 663, 844 

\bibitem[Whitmore et al.(1999)]{1999AJ....118.1551W} Whitmore, B.~C., 
Zhang, Q., Leitherer, C., Fall, S.~M., Schweizer, F., 
\& Miller, B.~W.\ 1999, \aj, 118, 1551 


\bibitem[Fryxell et al.(2000)]{2000ApJS..131..273F} Fryxell, B., et al.\ 
2000, \apjs, 131, 273 


\bibitem[Brandl et al.(2005)]{2005ApJ...635..280B} Brandl, B.~R., et al.\ 
2005, \apj, 635, 280 

\bibitem[Zhu et al.(2003)]{2003ApJ...588..243Z} Zhu, M., Seaquist, E.~R., 
\& Kuno, N.\ 2003, \apj, 588, 243 

\bibitem[Lin 
\& Murray(2007)]{2007ApJ...661..779L} Lin, D.~N.~C., \& Murray, S.~D.\ 2007, \apj, 661, 779 

\bibitem[Dull et al.(1997)]{1997ApJ...481..267D} Dull, J.~D., Cohn, H.~N., 
Lugger, P.~M., Murphy, B.~W., Seitzer, P.~O., Callanan, P.~J., Rutten, 
R.~G.~M., \& Charles, P.~A.\ 1997, \apj, 481, 267 


\bibitem[Whitmore et al.(2005)]{2005AJ....130.2104W} Whitmore, B.~C., et 
al.\ 2005, \aj, 130, 2104 

\bibitem[Pooley et al.(2003)]{2003ApJ...591L.131P} Pooley, D., et al.\ 
2003, \apjl, 591, L131 


\bibitem[Dinescu et al.(1997)]{1997AJ....114.1014D} Dinescu, D.~I., Girard, 
T.~M., van Altena, W.~F., Mendez, R.~A., 
\& Lopez, C.~E.\ 1997, \aj, 114, 1014 

\bibitem[Zezas et al.(2007)]{2007ApJ...661..135Z} Zezas, A., Fabbiano, G., 
Baldi, A., Schweizer, F., King, A.~R., Rots, A.~H., 
\& Ponman, T.~J.\ 2007, \apj, 661, 135 


\bibitem[Ruffert 
\& Arnett(1994)]{1994ApJ...427..351R} Ruffert, M., \& Arnett, D.\ 1994, \apj, 427, 351 

\bibitem[Smith et al.(1995)]{1995MNRAS.273..632S} Smith, G.~H., Woodsworth, 
A.~W., \& Hesser, J.~E.\ 1995, \mnras, 273, 632 

\bibitem[Knapp et al.(1996)]{1996ApJ...462..231K} Knapp, G.~R., Gunn, 
J.~E., Bowers, P.~F., \& Vasquez Poritz, J.~F.\ 1996, \apj, 462, 231 

\bibitem[Pfahl 
\& Rappaport(2001)]{2001ApJ...550..172P} Pfahl, E., \& Rappaport, S.\ 2001, \apj, 550, 172 

\bibitem[Kalogera et al.(2004)]{2004ApJ...601L.171K} Kalogera, V., King, 
A.~R., \& Rasio, F.~A.\ 2004, \apjl, 601, L171 

\bibitem[Maccarone et al.(2007)]{2007Natur.445..183M} Maccarone, T.~J., 
Kundu, A., Zepf, S.~E., \& Rhode, K.~L.\ 2007, \nat, 445, 183 


\bibitem[Menou et al.(1999)]{1999ApJ...520..276M} Menou, K., Esin, A.~A., 
Narayan, R., Garcia, M.~R., Lasota, J.-P., 
\& McClintock, J.~E.\ 1999, \apj, 520, 276 

\bibitem[Miller et al.(2003)]{2003ATel..212....1M} Miller, J.~M., Fabian, 
A.~C., \& Lewin, W.~H.~G.\ 2003, The Astronomer's Telegram, 212, 1 

\bibitem[Ulvestad et al.(2007)]{2007ApJ...661L.151U} Ulvestad, J.~S., 
Greene, J.~E., \& Ho, L.~C.\ 2007, \apjl, 661, L151 


\bibitem[Noyola et al.(2008)]{2008ApJ...676.1008N} Noyola, E., Gebhardt, 
K., \& Bergmann, M.\ 2008, \apj, 676, 1008 

\bibitem[Anderson 
\& van der Marel(2009)]{2009arXiv0905.0627A} Anderson, J., \& van der Marel, R.~P.\ 2009, arXiv:0905.0627 

\bibitem[Blondin 
\& Pope(2009)]{2009ApJ...700...95B} Blondin, J.~M., \& Pope, T.~C.\ 2009, \apj, 700, 95 


\bibitem[Portegies Zwart et al.(2004)]{2004Natur.428..724P} Portegies 
Zwart, S.~F., Baumgardt, H., Hut, P., Makino, J., 
\& McMillan, S.~L.~W.\ 2004, \nat, 428, 724 




\end{thebibliography}
%\bibliographystyle{apj}

\end{document}